\documentclass[11pt,twocolumns,article,nofontune]{IEEEtran}

\usepackage{url}
\usepackage[utf8]{inputenc}
\usepackage{xcolor}
\usepackage{amsmath}
\usepackage{amssymb}
\usepackage{xspace}
\usepackage{epsfig}
\usepackage{balance}

\usepackage[acronyms,nonumberlist,nopostdot,nomain,nogroupskip,acronymlists={hidden}]{glossaries}
\newglossary[algh]{hidden}{acrh}{acnh}{Hidden Acronyms}

\usepackage{booktabs}
\usepackage{tabularx}
\usepackage{multirow}

\usepackage[font=footnotesize]{subcaption}
\usepackage[font=footnotesize]{caption}

\usepackage{mathtools}
\usepackage[numbers,sort&compress]{natbib}

\usepackage{soul}

\newacronym{3gpp}{3GPP}{3rd Generation Partnership Project}
\newacronym{4g}{4G}{4th generation}
\newacronym{5g}{5G}{5th generation}
\newacronym{ai}{AI}{Artificial Intelligence}
\newacronym{cu}{CU}{Central Unit}
\newacronym{du}{DU}{Distributed Unit}
\newacronym{dl}{DL}{Deep Learning}
\newacronym{gnb}{gNB}{Next Generation Node Base}
\newacronym{mchem}{MCHEM}{Massive Channel Emulator}
\newacronym{mgen}{MGEN}{Multi-Generator}
\newacronym{ml}{ML}{Machine Learning}
\newacronym{mimo}{MIMO}{Multiple Input, Multiple Output}
\newacronym{mmwave}{mmWave}{millimeter wave}
\newacronym{pawr}{PAWR}{Platforms for Advanced Wireless Research}
\newacronym{powder}{POWDER}{Platform for Open Wireless Data-driven Experimental Research}
\newacronym{prb}{PRB}{Physical Resource Block}
\newacronym{qoe}{QoE}{Quality of Experience}
\newacronym{qos}{QoS}{Quality of Service}
\newacronym{ran}{RAN}{Radio Access Network}
\newacronym{rf}{RF}{Radio Frequency}
\newacronym{ru}{RU}{Radio Unit}
\newacronym{sla}{SLA}{Service Level Agreement}
\newacronym{srn}{SRN}{Standard Radio Node}
\newacronym{wg}{WG}{Working Group}
\newacronym{cnn}{CNN}{Convolutional Neural Network}
\newacronym{aoa}{AoA}{Angle of Arrival}
\newacronym[type=hidden]{nr}{NR}{New Radio}
\newacronym{lte}{LTE}{Long Term Evolution}
\newacronym{mac}{MAC}{Medium Access Control}
\newacronym{phy}{PHY}{Physical}
\newacronym{ric}{RIC}{\gls{ran} Intelligent Controller}
\newacronym{kpm}{KPM}{Key Performance Measurement}
\newacronym{mmtc}{mMTC}{Massive Machine-Type Communications}
\newacronym{mtc}{MTC}{Machine-type Communications}
\newacronym{urllc}{URLLC}{Ultra Reliable and Low Latency Communications}
\newacronym{embb}{eMBB}{Enhanced Mobile Broadband}
\newacronym{sdap}{SDAP}{Service Data Adaptation Protocol}
\newacronym{sm}{SM}{Service Model}
\newacronym{ei}{EI}{Enrichment Information}
\newacronym{smo}{SMO}{Service Management and Orchestration}
\newacronym{mcs}{MCS}{Modulation and Coding Scheme}
\newacronym{sinr}{SINR}{Signal to Interference plus Noise Ratio}
\newacronym{ue}{UE}{User Equipment}
\newacronym{drl}{DRL}{Deep Reinforcement Learning}
\newacronym{osc}{OSC}{O-RAN Software Community}
\newacronym{mns}{MnS}{Management Services}
\newacronym{pf}{PF}{Proportional Fair}
\newacronym{rr}{RR}{Round Robin}
\newacronym{rlc}{RLC}{Radio Link Control}
\newacronym{pdcp}{PDCP}{Packet Data Convergence Protocol}
\newacronym{srs}{SRS}{Sounding Reference Signal}

\glsdisablehyper

\definecolor{desireRed}{RGB}{230,57,60}%
\definecolor{darkPurple}{RGB}{59,31,43}%
\definecolor{springGreen}{RGB}{37,223,145}%
\definecolor{queenBlue}{RGB}{69,123,157}%
\definecolor{spaceCadet}{RGB}{29,53,87}%

\usepackage{dblfloatfix}

\IEEEoverridecommandlockouts

\begin{document}

\title{dApps: Distributed Applications for\\Real-time Inference and Control in O-RAN}

\author{\IEEEauthorblockN{Salvatore D'Oro, Michele Polese, Leonardo Bonati, Hai Cheng, Tommaso Melodia
\thanks{The authors are with the Institute for the Wireless Internet of Things, Northeastern University, Boston, MA, USA. E-mail: \{s.doro, m.polese, l.bonati, cheng.hai,  melodia\}@northeastern.edu}}
\thanks{This work was partially supported by the U.S.\ National Science Foundation under Grants CNS-1923789 and NSF CNS-1925601, and the U.S.\ Office of Naval Research under Grant N00014-20-1-2132.}}

\makeatletter
\patchcmd{\@maketitle}
  {\addvspace{0.5\baselineskip}\egroup}
  {\addvspace{-1.5\baselineskip}\egroup}
  {}
  {}
\makeatother

\flushbottom
\setlength{\parskip}{0ex plus0.1ex}

\maketitle
\glsunset{nr}
\glsunset{lte}
\glsunset{3gpp}
\glsunset{ran}  
\glsunset{phy}
\glsunset{mac}

\begin{abstract}
The Open \acrlong{ran} (Open \acrshort{ran})---being standardized, among others, by the O-RAN Alliance---brings a radical transformation to the cellular ecosystem through disaggregation and \glspl{ric} notions.
The latter enable 
closed-loop control through custom logic
applications, i.e., xApps and rApps, supporting control decisions at different timescales.
However, the current O-RAN specifications lack of a practical approach to execute \textit{real-time} control loops operating at timescales below 10\:ms.
In this paper, we propose the notion of {\em dApps}, distributed applications that complement existing xApps/rApps by allowing operators to implement fine-grained data-driven management and control in real-time at the \glspl{cu}/\glspl{du}.
dApps receive real-time data from the \gls{ran}, as well as enrichment information from the near-real-time \gls{ric}, and execute inference and control of lower-layer functionalities, thus enabling use cases with stricter timing requirements than those considered by the \glspl{ric}, such as beam management and user scheduling.
We propose feasible ways to integrate dApps in the O-RAN architecture by leveraging and extending interfaces and components already present therein.
Finally, we discuss challenges specific to dApps, and provide preliminary results that show the benefits of executing network intelligence through dApps.
\end{abstract}

\begin{picture}(0,0)(10,-420)
\put(0,0){
\put(0,10){\footnotesize This paper has been accepted for publication on IEEE Communications Magazine.}
\put(0,0){\tiny \copyright 2021 IEEE. Personal use of this material is permitted. Permission from IEEE must be obtained for all other uses, in any current or future media including reprinting/republishing}
\put(0,-6){\tiny this material for advertising or promotional purposes, creating new collective works, for resale or redistribution to servers or lists, or reuse of any copyrighted component of this work in other works.}}
\end{picture}

\glsresetall
\glsunset{nr}
\glsunset{lte}
\glsunset{3gpp}
\glsunset{ran}  
\glsunset{phy}
\glsunset{mac}

\vspace{-0.3cm}
\section{Introduction}

Cellular networks are undergoing a radical paradigm shift. One of the major drivers is the Open \acrlong{ran} (Open \acrshort{ran}) paradigm, which brings together concepts such as softwarization, disaggregation, open interfaces, and ``white-box" programmable hardware to supplant traditionally closed and inflexible architectures, thus laying the foundations for more agile, multi-vendor, data-driven, and optimized cellular networks~\cite{polese2022understanding}. 

\begin{figure}[t]
\setlength\belowcaptionskip{-0.7cm}
\begin{centering}
    \includegraphics[width=\columnwidth]{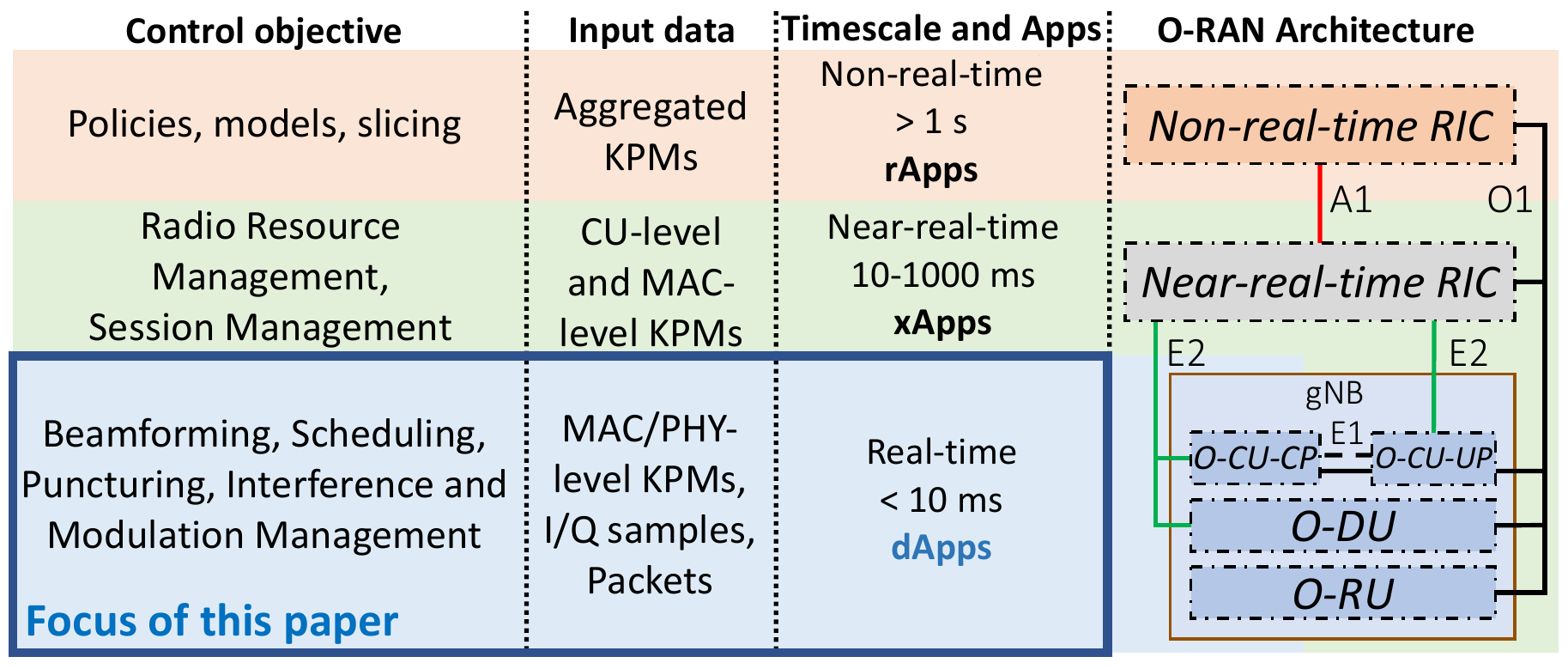}
    \caption{O-RAN architecture, applications and control loops.}
    \label{fig:table_apps}
\end{centering}
\end{figure}

This revolution is primarily led by the O-RAN Alliance, a consortium of network operators, vendors, and academic partners~\cite{polese2022understanding}.
O-RAN is standardizing the Open RAN architecture, its components and their functionalities, as well as
open interfaces to facilitate interoperability between multi-vendor components, real-time monitoring of the RAN, data collection and interactions with the cloud.
By adopting the~7.2x split, O-RAN builds upon the disaggregated 3GPP \glspl{gnb}, that divides the functionalities of the base stations across \glspl{cu}, \glspl{du}, and \glspl{ru}~\cite{polese2022understanding}.
As shown in Fig.~\ref{fig:table_apps}, O-RAN also introduces the concept of the \gls{ric}, an abstraction that enables near-real-time (or near-RT) and non-real-time (non-RT) control and monitoring of the RAN via software applications called xApps and rApps, respectively.
In the O-RAN vision, the components of the \gls{ran} expose a set of controllable parameters and functionalities, as well as streams of data (e.g., \glspl{kpm}). These are used by xApps and rApps to fine-tune the behavior of the \gls{ran}, adapting it to the operator goals, and to the network and traffic conditions through sophisticated \gls{ai} and \gls{ml} algorithms~\cite{bonati2021intelligence}.
The \glspl{ric}, xApps and rApps will eventually realize the vision of self-organizing networks autonomously detecting ongoing changes in channel, network, and traffic state, and reacting to meet minimum \gls{qos} requirements and to comply with \glspl{sla}. This includes resource allocation~\cite{zhang2022team}, network slicing~\cite{9490293}, handover and mobility management~\cite{orhan2021connection}, and spectrum coexistence~\cite{shi_rahimzadeh_costa_erpek_sagduyu_2021}.

However, we are still far from the vision of fully-automated and intelligent cellular networks.
Indeed, limiting the execution of control applications to the near-RT and non-RT \glspl{ric} prevents the use of data-driven solutions where control decisions and inference must be made in \textit{real time}, or within temporal windows shorter than the 10\:ms supported by near-RT control loops~\cite{bonati2021intelligence,abdalla2021toward}.
Two practical examples
are user scheduling and beam management.
Scheduling requires making decisions at sub-ms timescales (e.g., to perform puncturing and preemption to support \gls{urllc} traffic with latency values as low as 1\:ms).
Similarly, beam management
involves beam sweeping via reference signals transmitted within 5\:ms-long bursts (half the duration of a 5G NR frame)~\cite{polese2021deepbeam}.

Unfortunately, the near-RT \gls{ric} and xApps might struggle in accomplishing these procedures because they have limited access to low-level information (e.g., transmissions queues, I/Q samples, beam directionality) and/or incur high latency to obtain it.
For example, beam management would require the transmission of reference signals (or, as proposed in~\cite{polese2021deepbeam}, I/Q samples) from the \gls{du}/\gls{ru} to the \gls{ric} over the E2 interface.
This would result in increased overhead and delay due to propagation, transmission, switching, and inference latency, which might prevent real-time (i.e., $<\!\!\!10$\:ms) execution.
Moreover, since I/Q samples contain sensitive user data (e.g., packet payload), they cannot be transmitted to the \gls{ric} out of privacy and security concerns and are therefore processed at the gNB directly.
For these reasons, such procedures (and any procedure that requires real-time execution, or handles sensible data) are typically run directly at the \gls{du}/\gls{ru}, usually via closed and proprietary implementations---referred to as the ``vendor's secret sauce".
While hardware-based implementations can satisfy the above temporal requirements and deliver high performance, they are ultimately inflexible, hard to update, and not scalable as their upgrade (e.g., after a new 3GPP release) requires hardware or (whenever possible) firmware updates.

As of today, the O-RAN architecture focuses on offering softwarized, programmatic and \gls{ai}-based control to the higher layers of the protocol stack, with limited flexibility for the lower layers hosted at \glspl{du}/\glspl{ru}.
However, prior work has demonstrated how running \gls{ai} at the edge of the network---with a specific focus on \gls{phy} and \gls{mac} layers of the \glspl{du}/\glspl{ru}---can provide major performance benefits. Moreover, recent works have shown that AI at the edge can significantly improve network performance by leveraging traditionally available \glspl{kpm} (e.g., throughput, \gls{sinr}, channel quality information, latency)~\cite{9468707,9490293,9286741}, as well as by processing in parallel (thus not affecting demodulation and decoding procedures) I/Q samples collected at the PHY layer that carry detailed information on channel conditions and spatial information of received waveforms~\cite{polese2021deepbeam,oshea2017introduction}.
Although the O-RAN specifications have identified a few use cases that could benefit from running intelligence at \glspl{gnb} directly, these use cases are left for future studies~\cite{polese2022understanding}.

\noindent
\textbf{Main contributions.} The goal of this paper is to foster a discussion on enabling network intelligence at the edge in the O-RAN ecosystem.
As illustrated in Fig.~\ref{fig:table_apps}, we introduce the notion of \textit{dApps}, custom and distributed applications that complement xApps/rApps by implementing \gls{ran} intelligence at the \glspl{cu}/\glspl{du} for real-time use cases outside the timescales of the current \glspl{ric}.
dApps receive real-time data and \glspl{kpm} from the \glspl{ru} (e.g., frequency-domain I/Q samples), \glspl{du} (e.g., buffer size, \gls{qos} levels), and \glspl{cu} (e.g., mobility, radio link state), as well as \gls{ei} from the near-RT \gls{ric}, and use it to execute real-time inference and control of lower-layer functionalities.
We build on already available logical components and propose an extension of the O-RAN architecture to include the concept of dApps with minimal modification to the specifications.
Finally, we discuss challenges specific to dApps, and provide preliminary experimental results obtained on the Colosseum testbed~\cite{bonati2021intelligence,bonati2021colosseum} that demonstrate how dApps can enable a variety of real-time inference tasks at the edge and reduce control overhead.

\vspace{-0.3cm}
\section{Why dApps?}
\label{sec:why}

\textit{dApps} are \textit{distributed applications} that complement xApps/rApps to bring intelligence at \glspl{cu}/\glspl{du} and support real-time inference at tighter timescales than those of the \glspl{ric}. 
This section identifies their advantages and discusses relevant use cases and applications.
    
\noindent
\textbf{Reduced Latency and Overhead.}
Moving functionalities and services to the edge is one of the most efficient ways to reduce latency.
The near-RT \gls{ric} brings network control closer to the edge, but it
primarily executes in cloud facilities~\cite{polese2022understanding}.
Therefore, data still needs to travel from the \glspl{du} to 
the near-RT \gls{ric}, and the output of the inference needs to go back to the \glspl{du}/\glspl{ru}, which causes increased latency and overhead over the E2 interface to support data collection, inference and control. This can be mitigated by executing real-time procedures at the \glspl{cu}/\glspl{du} directly via dApps, which substantially reduces both latency and overhead (in Section~\ref{sec:usecases} we demonstrate a 3.57$\times$ overhead reduction).

\noindent
\textbf{AI at the Edge.}
While \gls{ai} (and specifically \gls{ml}) is usually associated with data centers with hundreds of GPUs,
nowadays there is plenty of evidence on
the feasibility of training and executing \gls{ai} on resource-constrained edge nodes with a limited footprint~\cite{jian2021radio}.
GPUs are now smaller, more powerful, cheaper, and widely available.
Technological advances in \gls{ai} have resulted in
procedures and techniques (e.g., pruning~\cite{jian2021radio}) that make it possible to compress \gls{ml}-solutions by 27$\times$ and reduce inference times by 17$\times$ while resulting in a negligible accuracy loss of 1\%.

\noindent
\textbf{Controlling MAC- and PHY-layer Functionalities.}
Another important aspect is related to controlling lower-layer functionalities of the \gls{mac} and \gls{phy} layers, such as
procedures related to scheduling, modulation, coding and beamforming, which all operate at sub-ms timescales and require real-time execution.
While xApps
can be used to select which scheduling policy to use at the \gls{du} (e.g., round-robin), they cannot allocate resource elements to \glspl{ue} in real time at the sub-frame level (e.g., to perform puncturing and preemption for \gls{urllc} traffic).
Moreover, many \gls{phy}-layer functionalities (e.g., beamforming, modulation recognition, channel equalization, radio-frequency fingerprinting-based authentication) operate in the I/Q domain and
recent advances show how those can be executed in software with increased flexibility, reduced complexity, and higher scalability by processing the I/Q samples directly~\cite{polese2021deepbeam}.
Because of these tight time constraints and security concerns,
xApps and rApps---which operate far from the \glspl{du}---unlike dApps, are not suitable to make decisions on these functionalities.

\noindent
\textbf{Access DU/CU Data and Functionalities in Real Time.}
dApps make it possible to access control- and user-plane data that is either unavailable at the near-RT \gls{ric}, or available but not with a sub-ms latency.
This includes real-time access to I/Q samples, data packets, handover-related mobility information,
dual-connectivity between 5G NR and 4G, among others.
By executing at the \glspl{du}/\glspl{cu}, dApps will be able to access \gls{ue}-specific metrics and data to deliver higher performance services tailored to individual \gls{ue} requirements, and instantaneous channel and network conditions. 

\noindent
\textbf{Extensibility and Reconfigurability.} 
Although there are rare cases where AI has been already embedded into \glspl{du} and \glspl{cu} the  majority of such solutions still leverage hardware-based implementations of \gls{mac} and \gls{phy} functionalities~\cite{polese2022understanding} that strongly limit their extensibility and reprogrammability.
On the contrary, the integration of dApps within the O-RAN ecosystem offers the ideal platform for software-based implementations of the above functionalities, and thus facilitates their  instantiation, execution and reconfiguration in real time and on demand. 
In this context, the O-RAN Alliance is developing standardized interfaces to support hardware acceleration in O-RAN~\cite{polese2022understanding}, which is a first step toward the integration of \gls{ai} within \glspl{du} and \glspl{ru}.

\vspace{-0.3cm}
\section{Challenges and Open Issues}
\label{sec:challenges}

Despite the above advantages, bringing intelligence to the edge comes with several challenges:

\noindent
\textbf{Resource management.} First, \gls{ai} solutions require computational capabilities to quickly and reliably perform inference. For this reason, the \glspl{du} must be equipped with enough computational power to support the execution of several concurrent dApps sharing the same physical resources without incurring in resource starvation and/or increased latency due to the instantiation and execution of many dApps on the same node. In this context, GPUs, CPUs, FPGAs, hardware acceleration and efficient resource virtualization, sharing and allocation schemes will play a vital role in the success of dApps.

\noindent
\textbf{Softwarized ecosystem.} Similar to the \gls{ric}, \glspl{cu}/\glspl{du} will need a container-based platform to support the seamless instantiation, execution, and lifecycle management of dApps. In contrast with other virtualization solutions (e.g., virtual machines), this offers a balanced trade-off between platform-independent deployment, portable and lightweight development and rapid instantiation and execution. At the same time, dApps must not halt or delay the real-time execution of gNB functionalities. In this context, hardware acceleration will be pivotal in guaranteeing that dApps execute reliably and fast.

\noindent
\textbf{Standardized interfaces for \glspl{du}/\glspl{cu}.} 
The execution of intelligence at the edge requires interfaces between \glspl{du}, \glspl{cu}, and dApps that offer similar functionalities to those currently available to the \glspl{ric} and other O-RAN components. 
This includes northbound (between dApps and the near-RT \gls{ric}) and southbound (between dApps and programmable functionalities and parameters of \glspl{du}/\glspl{cu}) interfaces. In this way, \glspl{du} can expose supported control and data collection capabilities to \glspl{cu} and the near-RT \gls{ric}. This is key to make sure that dApps are platform-independent and can seamlessly interact with other O-RAN components and applications.   

\noindent
\textbf{Orchestration of the intelligence.} dApps come with additional diversity and complexity.
This calls for
orchestration solutions that can determine which control and inference tasks are executed via dApps at \glspl{cu}/\glspl{du}, and which at the near-RT \gls{ric} via xApps according to data availability, control timescales, geographical requirements and network workload, while satisfying operator intents and \glspl{sla}. This also includes distributing network intelligence while avoiding conflicts between multiple O-RAN applications controlling RAN components.

\noindent
\textbf{Dataset availability.}
The reliability and robustness of AI for real-time inference and control will heavily rely upon availability of diverse and heterogeneous datasets. Large-scale Open RAN testbeds such as Colosseum~\cite{bonati2021colosseum} and digital twins will play a relevant role in generating those datasets and train, test and validate the effectiveness and generalization capabilities of dApps.

\noindent
\textbf{Friction from vendors.} Traditionally, gNB components host a large part of vendor's intellectual property (e.g., schedulers, beamforming, queue management). Enabling third-party applications at DUs and CUs will inevitably reduce the value of such intellectual property. Although the introduction of dApps may foster competitiveness and innovation, it might inevitably find friction from vendors. Another concern is often related to the monolithic development approach of \gls{ran} vendors, which would prevent the execution of third-party components such as dApps. Nonetheless, the xApp paradigm has already shown that it is possible to separate the \gls{ran} state machine between gNB nodes and the \glspl{ric} for control in the near- or non-real-time timescales. However, we would like to point out that these two aspects are not road blockers. Indeed, these have been already overcome in the historically closed market of networking solutions for data centers where, despite early frictions from manufacturers, Software Defined Networking (SDN) architectures and related solutions (e.g., P4, OpenFlow, Intel Tofino, to name a few) have taken over the market and demonstrated how real-time reprogrammability and open hardware are not only possible but extremely effective. This shows that monolithic, inflexible approaches are not the only option, and a similar approach to that of xApps/rApps can be adopted to implement dApps.

\begin{figure}[t]
\setlength\belowcaptionskip{-0.5cm}
\begin{centering}
    \includegraphics[width=\columnwidth]{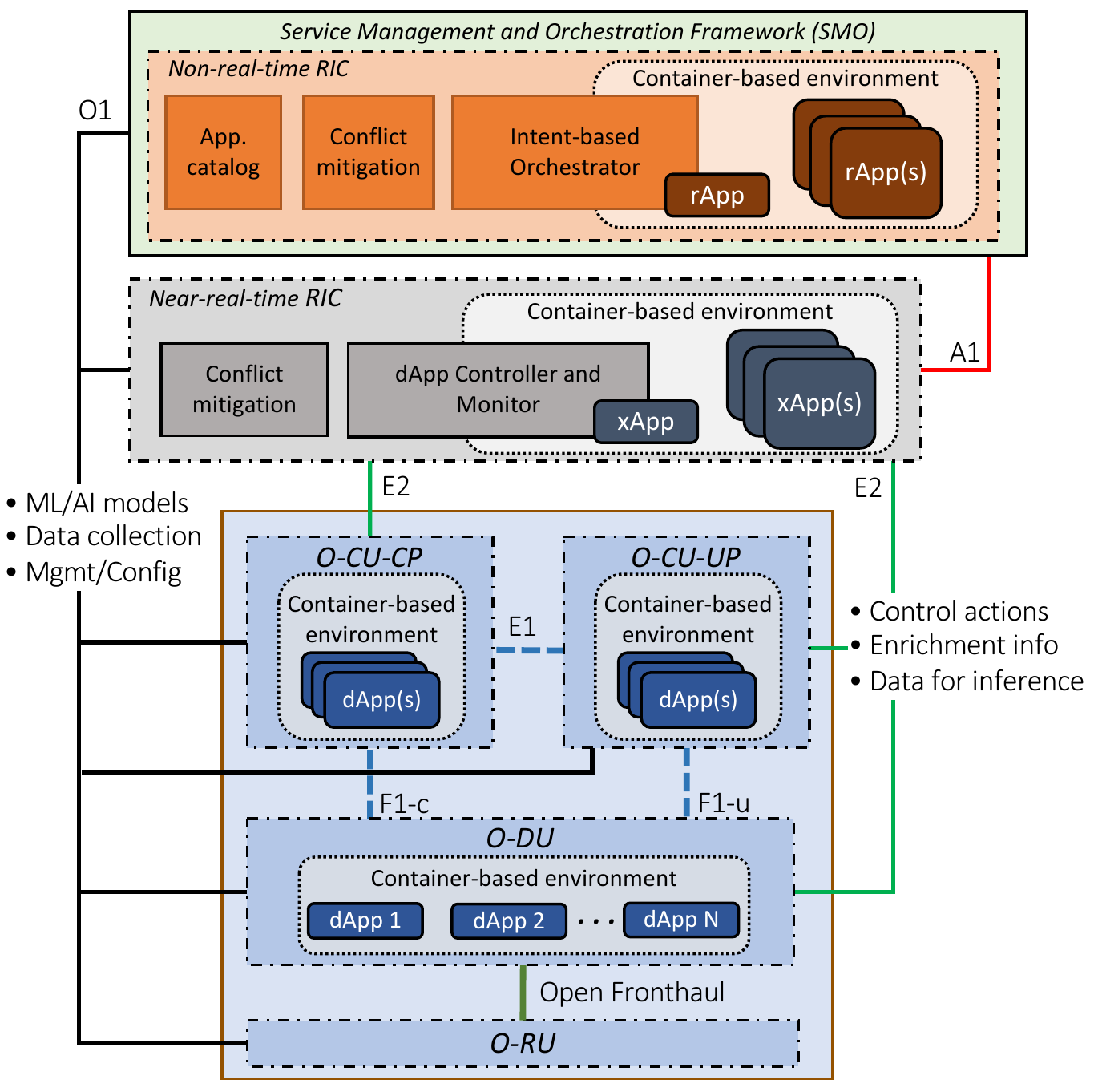}
    \caption{dApps and proposed extension to the O-RAN architecture.}
    \label{fig:architecture}
\end{centering}
\end{figure}

\vspace{-0.25cm}
\section{Proposed Architecture}
\label{sec:architecture}
\vspace{-0.1cm}

In this section, we discuss the architecture (shown in Fig.~\ref{fig:architecture}) necessary to support dApps while requiring minimal changes to the already existing O-RAN architecture.

\vspace{-0.45cm}
\subsection{dApps as Softwarized Containers}

Similarly to xApps and rApps, dApps leverage a containerized architecture to: (i)~seamlessly manage the lifecycle of dApps, i.e., deployment, execution and termination; (ii)~facilitate the integration and use of new (or updated) functionalities included in newly-released O-RAN specifications via software updates; (iii)~provide an abstraction level where the
\glspl{cu}, \glspl{du}, and \glspl{ru}
advertise the tunable parameters and functionalities (similarly to what is already envisioned for xApps and the E2 interface) to enable dApps tailored to control specific parameters; (iv)~achieve hardware-independent implementations of dApps, which can be offered as standalone O-RAN applications in a marketplace that fosters innovation and competition via openness, and (v)~facilitate the development and use of \gls{ai}-based solutions for the lower layers of the protocol stack.
This approach also requires a resource manager in place that allows containers to access and share the physical resources (e.g., CPUs, GPUs, memory) available in the \gls{ran} nodes.

\vspace{-0.5cm}
\subsection{Leveraging O-RAN Interfaces}

The O-RAN interfaces currently available can be extended and used to support the deployment, execution and management of dApps:

\noindent
\textbf{Southbound Interfaces.} Currently, the O-RAN specifications do not envision
data-driven control based on analysis and inference of user-plane data, including I/Q samples and data packets. These, however, can be the basis for several data-driven use cases, discussed in Section~\ref{sec:usecases}. To support these use cases, dApps require southbound interfaces to allow dApps executing at the \gls{du} to receive (i) waveform samples in the frequency domain from the \gls{ru} over the O-RAN Fronthaul interface,
as well as (ii) transport blocks, or \gls{rlc} packets that are already locally available at the \gls{du}. 
Similarly, southbound interfaces must allow dApps executing at the \gls{cu} to perform inference on locally available data pertaining to \gls{pdcp} and \gls{sdap}.
As of today, these southbound interfaces are not yet available, but we propose to implement them by adapting and extending the \glspl{sm} defined for the E2 interface. In this way, dApps can extract relevant \glspl{kpm} using the southbound E2-like \gls{sm} \gls{kpm} adapted to support dApps within a latency of 10\:ms to support real-time execution.

\noindent
\textbf{Northbound Interfaces.} Similar to how xApps receive \gls{ei} from the non-RT \gls{ric} via the A1 interface, dApps can receive \gls{ei} from the near-RT \gls{ric} via the E2 interface.
In this case, xApps
process data from one or more gNBs, and send \gls{ei} to the dApps, which use it to make decisions on control operations. For example, a \gls{du} can receive traffic forecasts from the near-RT \gls{ric}, and use this information to control scheduling, \gls{mcs}, and beamforming. 
Similarly to xApps, dApps are dispatched via the O1 interface.

\vspace{-0.5cm}
\subsection{Extending Conflict Mitigation to dApps}

The O-RAN specifications envision conflict mitigation components to ensure that the same parameter or functionality (e.g., scheduling policy of a gNB) is controlled by at most one O-RAN application at any given time.
The introduction of dApps
will further emphasize the importance of conflict detection and mitigation at stricter timescales than those currently envisioned by O-RAN.
Indeed, dApps require conflict mitigation to identify conflicts between rApps, xApps and dApps. In this context, pre-action conflict resolution (such as those envisioned for the near-RT RIC~\cite{polese2022understanding}) can prevent directly observable conflicts between different applications (e.g., two applications controlling the same parameter). On the contrary, those conflicts that cannot be observed directly, i.e., implicit conflicts where two or more applications control different parameters indirectly affecting the same set of KPMs, can be mitigated through post-action verification where conflicts are detected by observing the impact and extent that control actions taken by different O-RAN applications have on the same KPMs.

\vspace{-0.5cm}
\subsection{Intent-based O-RAN Apps Orchestrator}

The abundance of O-RAN applications will require automated solutions capable of determining which applications should be executed and where.
This task is left to the orchestration module shown in Fig.~\ref{fig:architecture} residing in the non-RT \gls{ric}
and executing either as an rApp, or as a standalone component within the \gls{smo} domain.
This module converts goals and requirements of the operator (e.g., in YAML/XML/JSON format) into a set of O-RAN applications that constitute a fabric of intelligent modules embedding the necessary AI to meet the desired intent.
Then, it dispatches them from the application catalog where they reside to the \gls{ran} location where they are executed, thus creating a complex ecosystem of applications that cooperate to achieve the operator intent.

To achieve this, the orchestrator needs to understand the intent specified by the operator, and compute the optimal configuration and set of applications to instantiate and where~\cite{doro2022orchestran}.
This is performed by ensuring that applications are executed only at network nodes: (i)~where input data can be made available within the required timescale; (ii)~that can actually control the required parameters and functionalities, and (iii)~with enough physical resources (e.g., CPUs/GPUs/FPGAs) to support the required applications.
For example, if an operator wants to perform real-time beam detection and traffic forecasting for a set of gNBs,
the orchestrator needs to deploy a dApp that executes at the \gls{du} (where the I/Q samples are available through the Open Fronthaul interface) to perform beam detection, and an xApp at the near-RT \gls{ric} (that receives traffic-related \glspl{kpm} from the \glspl{cu} via the E2 interface) to perform traffic forecasting.

\vspace{-0.5cm}
\subsection{dApp Controller and Monitor}

This component is hosted in the near-RT \gls{ric}, (Fig.~\ref{fig:architecture}) and is in charge of controlling and monitoring dApps executing at the gNBs.
Specifically, it ensures that dApps meet the desired \gls{qos} levels and are in line with the operator intent. As a possible extension, this component can also
convert an xApp into multiple atomic dApps dispatched and executed at the gNB components to provide a
finer control of the \gls{ran} procedures. In this case, the dispatchment
can be coordinated by the non-RT \gls{ric}, and performed via the O1 interface.

\vspace{-0.2cm}
\section{Use Cases and Results}
\label{sec:usecases}

Now we discuss relevant use cases that would benefit from dApps and present preliminary results that demonstrate how dApps can effectively reduce overhead over O-RAN interfaces while supporting AI solutions for real-time control of the RAN.

\vspace{-0.5cm}
\subsection{Beam Management}

dApps can be used to extend the beam management capabilities of NR \glspl{gnb}. 
The 3GPP
specifies a set of synchronization and reference signals to evaluate the quality of specific beams, and to allow the \gls{ue} and the \gls{ran} to use intelligent algorithms~\cite{polese2021deepbeam,zhou2017beam} that select the best combination of transmit and receive beams. 
These techniques, however, require a dedicated implementation on \gls{ran} components that vendors offer as a black box. In this case, xApps and rApps can only embed logic to control high-level parameters, e.g., select and deploy a codebook at the \gls{ru} based on \glspl{kpm} or coarse channel measurements.
On the contrary, dApps can support custom beam management logic where the dApp itself selects the beams to use and/or explore, rather than xApps providing high-level policy guidance.

For example, in~\cite{polese2021deepbeam} we introduced DeepBeam, a beam management framework that leverages deep learning on the I/Q samples to infer the \gls{aoa} and which beam is the transmitter using in a certain codebook. 
DeepBeam is thus an example of a data-driven algorithm that cannot be deployed at the
\glspl{ric}, as it requires access to user-plane I/Q samples for inference. This approach is an ideal candidate for deployment in a dApp, as it requires access to information that can be easily exposed by a \gls{du} in real time (i.e., the frequency-domain waveform samples), but cannot be transferred to another component of the network without (i)~violating control latency constraints, (ii) exposing sensitive user data; and (iii)~increasing the traffic on the E2 or O1 interface excessively.
\begin{figure}
\setlength\belowcaptionskip{-0.3cm}
    \centering
    \includegraphics[width=\columnwidth]{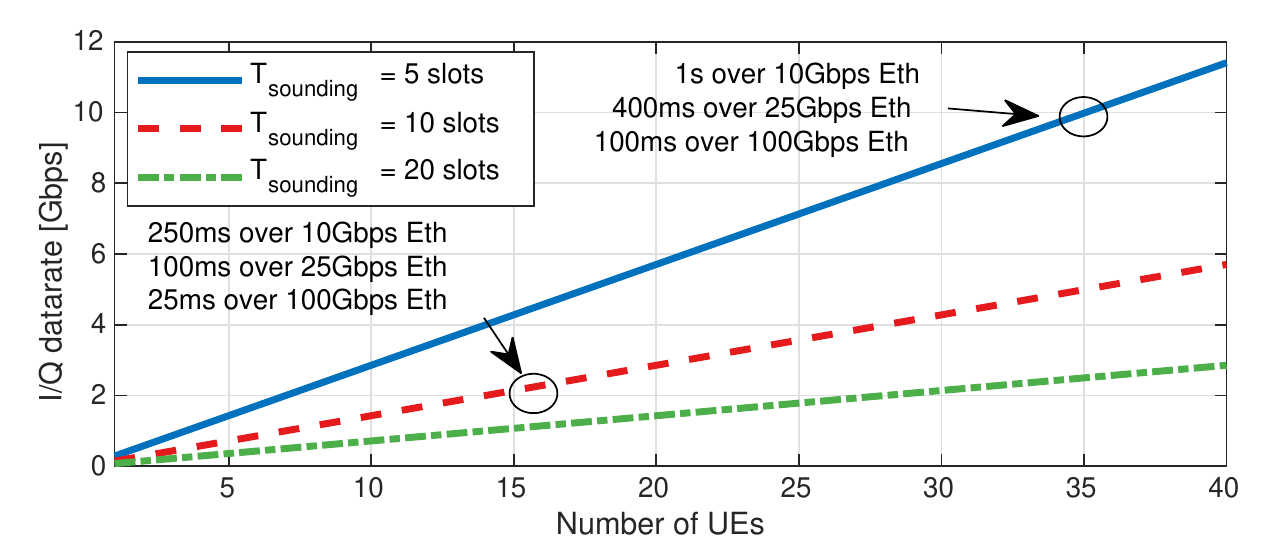}
    \caption{Data rate and latency to perform I/Q-based beam management over E2 interface. In most cases, latency is higher than 10\:ms (required to perform real-time beam management).}
    \label{fig:deepbeam}
\end{figure}
As an example, Fig.~\ref{fig:deepbeam} reports the data rate (and time) needed to transfer the I/Q samples required to perform inference with the DeepBeam convolutional neural networks from a \gls{du} to the near-RT \gls{ric}. DeepBeam can perform inference and classify the transmit beam and \gls{aoa} using any kind of samples (e.g., from packets or sounding signals)~\cite{polese2021deepbeam}. As a reference, in this case we consider the number of samples that can be collected through 3GPP NR \glspl{srs}. We use 3GPP-based parameters and assume that each \gls{srs} uses 3300 subcarriers (i.e., the full bandwidth available to NR \glspl{ue}), 2 symbols in time, a periodicity $T_{\rm sounding}$ of 5, 10, or 20 slots, and that each \gls{ue} monitors 3 uplink beams. The I/Q samples have 9 bits, and we assume numerology 3 (i.e., slots of $125\:\mathrm{\mu s}$). The results show that it would be impractical to transfer the required amount of samples because of timing (i.e., no real-time control) and of the data rate required, which can reach more than 100 Gbps in certain configurations.

\vspace{-0.3cm}
\subsection{Supporting Low-latency Applications}

Another application of practical relevance is that of dApps to support real-time and low-latency applications by, for example, controlling RAN slicing and scheduling decisions. 
Indeed, the timescale at which dApps operate is appropriate to access \gls{ue}-specific information from the \gls{du} in real time (e.g., buffer size, \gls{mcs} profile, instantaneous \gls{sinr}), and to make decisions on the RAN slicing and resource allocation strategies based on \gls{qos} requirements and network conditions.

\begin{figure}[t]
\setlength\belowcaptionskip{-1.3cm}
    \begin{minipage}[t]{0.37\columnwidth}
    \centering
    \includegraphics[width=\columnwidth]{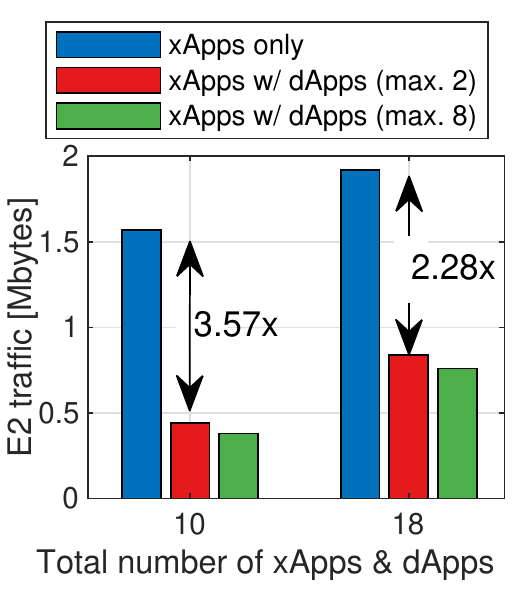}
    \caption{\label{fig:e2_schedl_slicing}E2 traffic analysis.}
    \vspace{4ex}
  \end{minipage}
  \hspace{0.1cm}
\begin{minipage}[t]{0.62\columnwidth}
    \centering
    \includegraphics[width=\columnwidth]{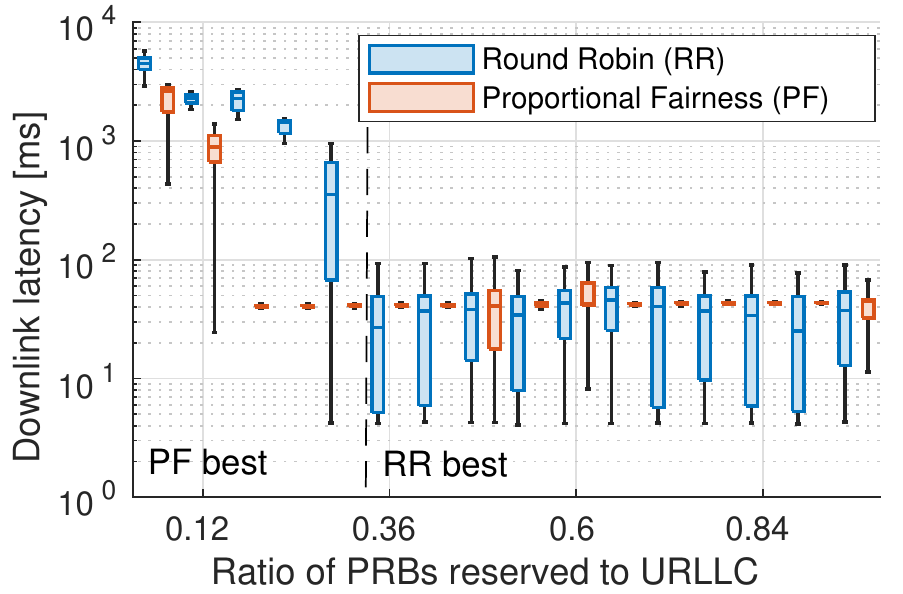}
    \caption{\label{fig:latency_urllc}URLLC slice end-to-end latency for different RAN slicing and schedulers.}
    \vspace{4ex}
  \end{minipage} 
\end{figure}

To showcase the benefits of dApps, we trained a set of \gls{ml} solutions for O-RAN applications. Specifically, we trained two \gls{drl} agents that process input data from the \gls{ran} (i.e., downlink buffer occupancy, throughput, traffic demand) to control the scheduling and RAN slicing policies of the gNBs (due to space limitations, details are omitted and can be found in~\cite{bonati2021intelligence}).
The gNBs are deployed on the Colosseum platform~\cite{bonati2021colosseum} ands implement network slices associated to different traffic types, i.e., \gls{embb}, \gls{mtc}, and \gls{urllc} traffic.
The agents aim at (i)~maximizing the throughput for the \gls{embb} slice, (ii)~maximizing the number of transmitted packet for \gls{mtc}, and (iii)~reducing the service latency for \gls{urllc}. Moreover, we also trained two forecasting models to predict the \gls{ue} traffic demand and the transmission buffer occupancy.

We consider the case where the \gls{drl} agents and the forecasters can run either at the near-RT \gls{ric} as xApps, or at the \glspl{du} as dApps. Both xApps and dApps have been implemented as Docker containers. In the former case, data for inference is received from the E2 interface, while in the latter data is locally available at the dApp.
We also leverage the OrchestRAN~\cite{doro2022orchestran} framework developed in our prior work to orchestrate the network intelligence according to operator's intents, determine how to split and distribute intelligence among xApps and dApps, and dispatch them.
Figure~\ref{fig:e2_schedl_slicing} shows the impact that running intelligence at the dApps has on the overhead over the E2 interface as a function of the total number of deployed xApps and dApps.
We consider three different configurations. In one configuration, the intelligence can only run at the xApps; in the other two, the ML solutions can be executed either through xApps or dApps, with the \glspl{du} supporting at most 2 and 8 concurrent dApps.
Figure~\ref{fig:e2_schedl_slicing} shows that dApps halve the traffic over the E2 interface, with a traffic reduction up to 3.57$\times$ with respect to the case with only xApps. Notice that two or more xApps can share the same input data received over the E2 interface. Thus, the traffic over E2 does not linearly grow with the number of xApps.

To further demonstrate the importance of controlling RAN behavior in real time, we ran extensive data collection campaigns on Colosseum~\cite{bonati2021intelligence,bonati2021colosseum}, and demonstrated the impact of selecting different RAN slicing (i.e., the ratio of \glspl{prb} reserved exclusively to \gls{urllc} traffic) and scheduling strategies (i.e., \gls{rr} and \gls{pf}) on the application-layer latency of URLLC traffic.
The results reported in Fig.~\ref{fig:latency_urllc} demonstrate the importance of joint slicing and scheduling control to support URLLC use cases. For example, when less than 30\% of resources are reserved for the URLLC traffic, selecting the \gls{pf} scheduling algorithm ensures the lowest latency. On the contrary, \gls{rr} works best when more \glspl{prb} are reserved to URLLC communications, with end-to-end latency values as low as 4\:ms.
These results show that achieving ultra-low latency still requires decisions made at the \glspl{du} directly via dApps to ensure a tolerable end-to-end latency level despite rapidly changing channel and network conditions (e.g., buffer size, traffic load).

\vspace{-0.2cm}
\section{Conclusions}
\label{sec:conclusions}

The availability of data-driven, custom control logic is one of the major benefits of the O-RAN architecture. 
In this paper, we proposed to extend this even further with the concept of dApp, distributed O-RAN applications executing at the \gls{du} and \gls{cu}, and complementing xApps and rApps. We first discussed the benefits introduced by dApps, which include real-time control for a set of parameters that cannot otherwise be optimized with near-RT or non-RT control loops. We discussed challenges, related to standardization, the need for resources and softwarized platforms, and orchestration of the functionalities. We then provided details on an architectural extension that enables the dApp vision, and illustrated two relevant use cases. 
Our discussion shows that while dApps are well-suited to augment O-RAN control and monitoring operations, there are still a few aspects that need further analysis. These include integration with data factories and digital twins for reliable AI, well-defined interfaces between dApps and CU/DU functionalities, and reduce frictions from vendors.

\balance
\footnotesize  
\bibliographystyle{IEEEtran}
\bibliography{biblio.bib}

\vspace{-1.3cm}
\begin{IEEEbiographynophoto}
{Salvatore D'Oro} [M'17] is a Research Assistant Professor with Northeastern University. He received his Ph.D.\ from the University of Catania in 2015. His research focuses on optimization and learning for NextG systems and the Open RAN.
\end{IEEEbiographynophoto}

\vspace{-1.3cm}
\begin{IEEEbiographynophoto}{Michele Polese}
[M'20] is a research scientist at Northeastern University. He obtained his Ph.D.\ from the University of Padova, Italy, in 2020, where he also was a postdoctoral researcher and adjunct professor. His research focuses on architectures for wireless networks.
\end{IEEEbiographynophoto}

\vspace{-1.3cm}
\begin{IEEEbiographynophoto}{Leonardo Bonati} [S'19] is a Ph.D.\ candidate at Northeastern University. He received his M.S.\ in Telecommunication Engineering from the University of Padova, Italy in 2016. His research focuses on softwarized NextG systems.
\end{IEEEbiographynophoto}

\vspace{-1.3cm}
\begin{IEEEbiographynophoto}{Hai Cheng} is a Ph.D.\ candidate in Computer Engineering at the Institute for Wireless IoT at Northeastern University. His research interests include machine learning and optimization in wireless network systems.
\end{IEEEbiographynophoto}

\vspace{-1.3cm}
\begin{IEEEbiographynophoto}{Tommaso Melodia}
[F’18] received a Ph.D.\ in Electrical and Computer Engineering from the Georgia Institute of Technology in 2007. He is the William Lincoln Smith Professor at Northeastern University, the Director of the Institute for the Wireless Internet of Things, and the Director of Research for the PAWR Project Office. His research focuses on wireless networked systems.
\end{IEEEbiographynophoto}

\end{document}